\documentclass[
twocolumn,
 amsmath,amssymb,
 aps,
prb,
]{revtex4-2}

\usepackage{graphicx}
\usepackage{dcolumn}
\usepackage{bm}
\usepackage{subfigure}
\usepackage{upgreek}
\usepackage{hyperref}
\usepackage{ulem}

\usepackage{xcolor}

\begin{document}

\preprint{APS/123-QED}

\title{Metallic conductivity on Na-deficient structural domain walls in the spin-orbit Mott insulator Na$_2$IrO$_3$}

\author{Franziska A. Breitner}
 \email{franziska.breitner@physik.uni-augsburg.de}

\author{Julian Kaiser}
\author{Anton Jesche}
\author{Philipp Gegenwart}
 \email{philipp.gegenwart@physik.uni-augsburg.de}
\affiliation{
Experimental Physics VI, Center for Electronic Correlations and Magnetism, University of Augsburg, Universitätsstr. 1, 86159 Augsburg, Germany}

\date{\today}

\begin{abstract}
Honeycomb Na$_2$IrO$_3$ is a prototype spin-orbit Mott insulator and Kitaev magnet. We report a combined structural and electrical resistivity study of Na$_2$IrO$_3$ single crystals. Laue back-scattering diffraction indicates twinning with $\pm 120^\circ$ rotation around the $c^*$-axis while scanning electron microscopy displays nanothin lines parallel to all three b-axis orientations of twin domains. Energy dispersive x-ray analysis line-scans across such domain walls indicate no change of the Ir signal intensity, i.e. intact honeycomb layers, while the Na intensity is reduced down to $\sim 2/3$ of its original value at the domain walls, implying significant hole doping. Utilizing focused-ion-beam micro-sectioning, the temperature dependence of the electrical resistance of individual domain walls is studied. It demonstrates the tuning through the metal-insulator transition into a correlated-metal ground state by increasing hole doping.


\end{abstract}

\maketitle


\section{{Introduction}}

Spin-orbit (SO) coupling describes the interaction between the spin and the orbital motion of electrons. In condensed matter composed of light elements, SO coupling is very weak and often neglected. However, for heavy atoms with high nuclear charge, a substantial magnetic field in the reference frame of moving electrons couples to their spins and SO coupling is the driving source for novel quantum states, such as topologically protected conducting surface states in topological insulators (TIs)~\cite{More}. Established TIs, based on semiconductors with s- and p-orbitals, are well described by the band theory of non-interacting electrons. Challenging for theory and experiment are various topological phases in highly correlated electron systems~\cite{pesin_mott_2010,Dzero2010,Rau2016,Tokura2022}.

Iridates are promising in this respect: due to the large atomic mass of iridium ions, a strong SO coupling of order 0.4 eV has been found~\cite{kim_novel_2008,watanabe_microscopic_2010,Clancy2012}, reaching the typical size of the Coulomb repulsion for 5d electrons. For the 5d$^5$ electrons of Ir$^{4+}$ ions in an octahedral crystal electric field the strong SO coupling breaks the sixfold degeneracy of the t$_{\rm 2g}$ states leading to filled $j_{\rm eff} = 3/2$ and half-filled $j_{\rm eff} = 1/2$ bands. Since the latter band is narrow, even a small Coulomb repulsion could then open a Mott gap leading to a $j_{\rm eff} = 1/2$ SO Mott insulator. The corresponding wave function is different from that of ordinary $s=1/2$ Mott insulators as it contains an equal mixture of xy, yz and zx orbitals with mixed spin up and down states~\cite{kim_novel_2008}. One of the wave function’s components is purely complex. Thus, 5d electrons will acquire a complex phase upon hopping via oxygen p-orbitals. The experimental proof for the complex $j_{\rm eff} = 1/2$ SO Mott state has been obtained by resonant inelastic x-ray scattering~\cite{kim_phase-sensitive_2009}. The complex transfer integral for the hopping could lead, in suitable crystal structures, to topologically nontrivial states. Such scenarios have been discussed for pyrochlore and honeycomb iridates~\cite{pesin_mott_2010,shitade_quantum_2009,Rau2016}. 
The $j_{\rm eff} = 1/2$ spins in Na$_2$IrO$_3$ also give rise to a bond-dependent magnetic exchange that can be described by an extended Kitaev model~\cite{chaloupka_kitaev-heisenberg_2010}. Though these moments display a zigzag antiferromagnetic order at $\sim 15$~K ~\cite{singh_antiferromagnetic_2010,choi_spin_2012,ye_direct_2012}, there is clear evidence for the presence of a dominating Kitaev interaction~\cite{singh_relevance_2012,hwan_chun_direct_2015,winter_models_2017,revelli_fingerprints_2020} at elevated temperatures and excitation energy.

Tuning SO Mott insulators by charge carrier doping into the metallic state seems interesting and promising. It is motivated, for instance, by the analogy between the perovskite $j_{\rm eff} = 1/2$ SO Mott insulator Sr$_2$IrO$_4$ and the isostructural $s=1/2$ Mott insulator La$_2$CuO$_4$, parent compound for high-$T_c$ superconductivity. While angle-resolved photoemission spectroscopy (ARPES)  and scanning tunneling microscopy (STM) indeed recorded signatures of a pseudogap in electron-doped (Sr$_{1-x}$La$_x$)$_2$IrO$_4$~\cite{Kim2016,Battisti2017}, a fully metallic or superconducting electrical resistance behavior has not yet been reported for doped or pressurized Sr$_2$IrO$_4$.  The $j_{\rm eff} = 1/2$ SO Mott state in honeycomb Na$_2$IrO$_3$ has been confirmed by the analysis of resonant inelastic x-ray scattering spectra~\cite{Gretarsson2013} with charge gap of 0.34 eV, determined by ARPES and optical conductivity~\cite{comin_na2iro3_2012}. While Shitade {\it et al.} treated Na$_2$IrO$_3$ in the weak correlation limit as layered quantum spin Hall insulator~\cite{shitade_quantum_2009}, it is now clear, that strong correlations are needed to account for its magnetism~\cite{winter_models_2017} and the size of the charge gap~\cite{comin_na2iro3_2012}. In the absence of a clear hierarchy of the energy scales for SO coupling, Coulomb repulsion, one-electron hopping and trigonal crystal field splitting, the band structure of Na$_2$IrO$_3$ has been proposed to sensitively depend on small structural variations and quantum phase transitions, separating trivial normal, metallic and topological phases  as function of hopping strength were proposed earlier~\cite{Choong2012}. Even the possibility of topological spin-triplet superconductivity in charge carrier doped honeycomb iridates was considered theoretically~\cite{you_doping_2012}.

While the bulk electrical resistance of Na$_2$IrO$_3$ displays three-dimensional variable-range hopping of carriers localized by disorder~\cite{singh_antiferromagnetic_2010,jenderka2013,rodriguez_competition_2020}, similar as in other iridates, different experiments suggested the possibility of enhancing the conductivity at the surface. An ARPES study reported an in-gap feature hinting at a metallic surface state~\cite{alidoust_observation_2016}, that was later attributed to dispersive
states approaching the Fermi level in the Na terminated cleavage planes~\cite{moreschini_quasiparticles_2017}. Indeed atomic resolution STM found different Na-deficient surface reconstructions~\cite{lupke2015}. Careful analysis of tunneling spectra at differently reconstructed surfaces revealed defect states in the bulk gap, as well as a V-shaped bandgap closing that was associated to surface states~\cite{dziuba_combined_2022}. A reduced and temperature independent electrical resitivity, measured under UHV conditions on a freshly cleaved crystal, was related to such a surface conductivity channel~\cite{dziuba_combined_2022}. On the other hand, argon plasma etching the surface of Na$_2$IrO$_3$  crystals after cleaving in air revealed first-order density-wave-like transitions in the electrical resistance around $\sim 220$~K and a Fermi liquid behavior below 20~K~\cite{mehlawat2016}.

Motivated by these hints at a high tunability of the electronic properties we decided to perform a combined structural and electrical resistivity study on Na$_2$IrO$_3$. We started with carefully searching for possible structural domains. Indeed, Laue backscattering diffraction indicates three different orientations of the honeycomb lattice with mutual $\pm 120^\circ$ rotations. Interestingly, these domain boundaries are seen as straight long white lines by scanning electron microscopy and feature a significant Na deficiency. Focused-ion-beam (FIB) micro-sectioning of the domain walls allowed us to study their electrical resistance behavior. We find a clear tendency that increasing Na deficiency reduces the resistance and finally induces a completely metallic state down to lowest temperatures, indicating that Na$_2$IrO$_3$ has been tuned locally across the metal-insulator transition by hole doping.



\section{{Methods}}

High quality single crystals of Na$_2$IrO$_3$ of several mm diameter were grown via solid state reaction as described in~\cite{singh_antiferromagnetic_2010}. For Laue back-scattering measurements a Philips PW1830/40 x-ray generator with tungsten anode and Photonic Science detector were utilized. The diffraction patterns were recorded with voltages of 10, 15 and 20\,kV and a current of 30\,mA. An exposure time of 10 minutes was deemed sufficient, as prolonged exposure did not yield significant improvement of the diffraction patterns for crystals of good quality. Attempts were made to further improve the quality of the measured Laue patterns by polishing or cleaving the samples. However, no significant improvement was achieved. Data analysis of the Laue patterns was performed using the Crystal Maker software Single Crystal as well as our own program which allows to consider additive overlap of reflections as well as reflections due to proximity to the characteristic wavelengths of tungsten. A scanning electron microscope (SEM) Merlin Gemini 2 from Zeiss, equipped
with an energy dispersive x-ray analysis (EDX) probe with X-Max 80N SDD detector from Oxford Instruments was utilized for structural and compositional investigation of cleaved Na$_2$IrO$_3$ surfaces. Silver epoxy was used to mount the crystals onto the sample holder. The dual beam FIB-SEM Crossbeam 550 from Zeiss was used for microstructuring
and electrically contacting small surface areas for electrical resistivity measurements. The measurements were performed between 2 and 300~K in the Dynacool PPMS with ETO option from Quantum Design. All measurements were done with an excitation current of $3\cdot 10^{-5}$\,mA. 

\section{{Results and Discussion}}

\subsection{Structural domains in Na$_2$IrO$_3$}

\begin{figure}[ht]
\subfigure[]{\includegraphics[width=0.42\linewidth]{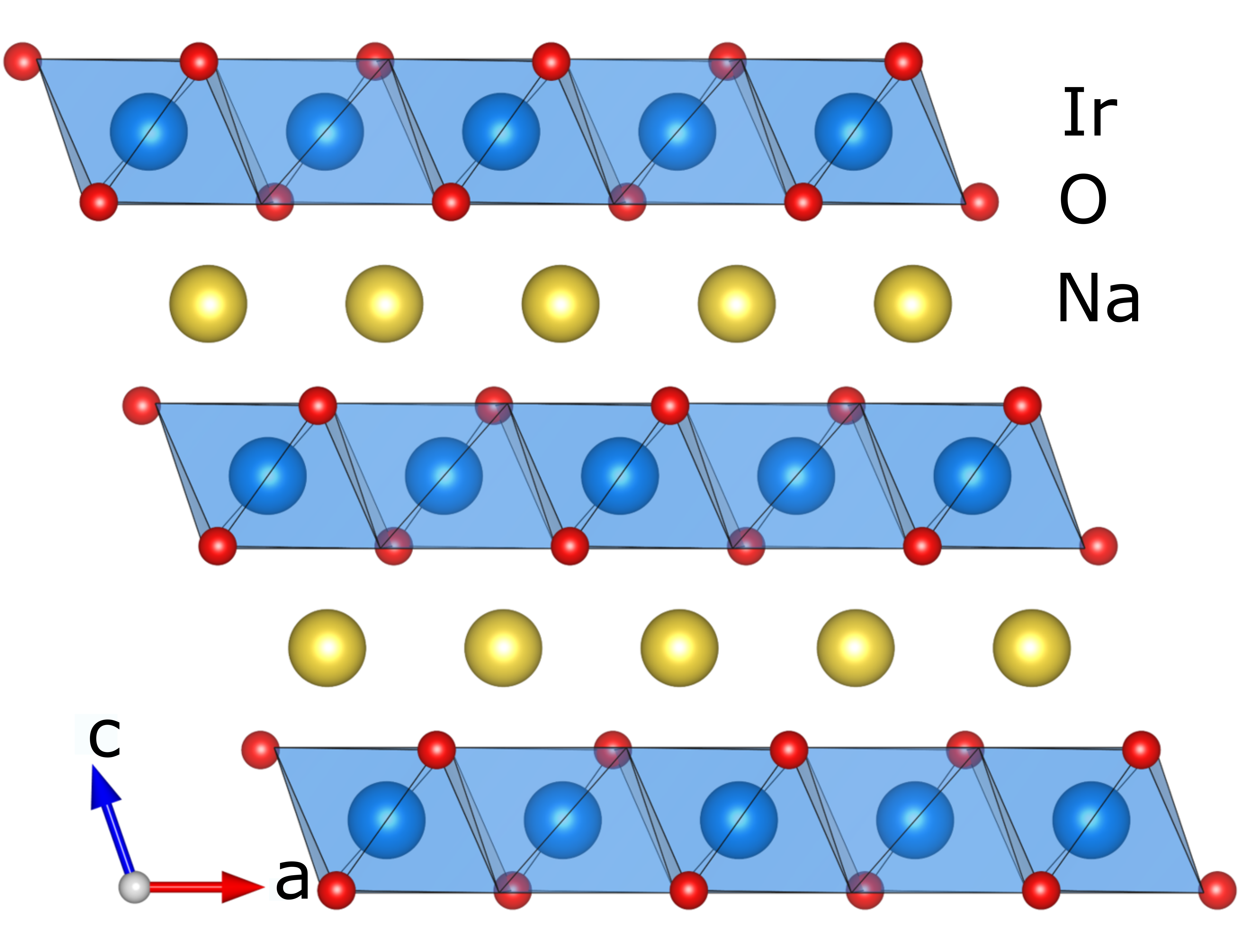}}
\subfigure[]{\includegraphics[width=0.261\linewidth]{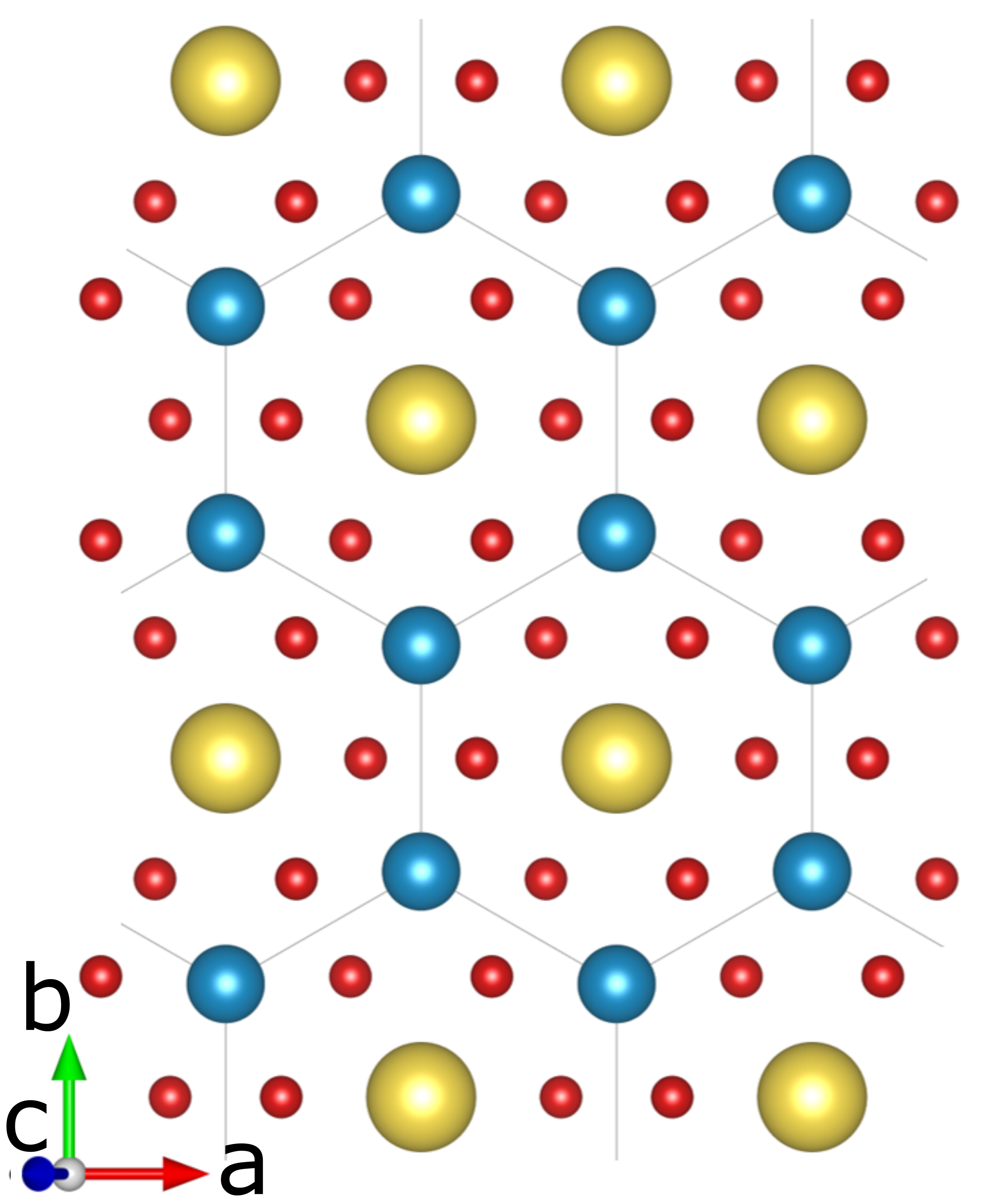}}
\subfigure[]{\includegraphics[width=0.29\linewidth]{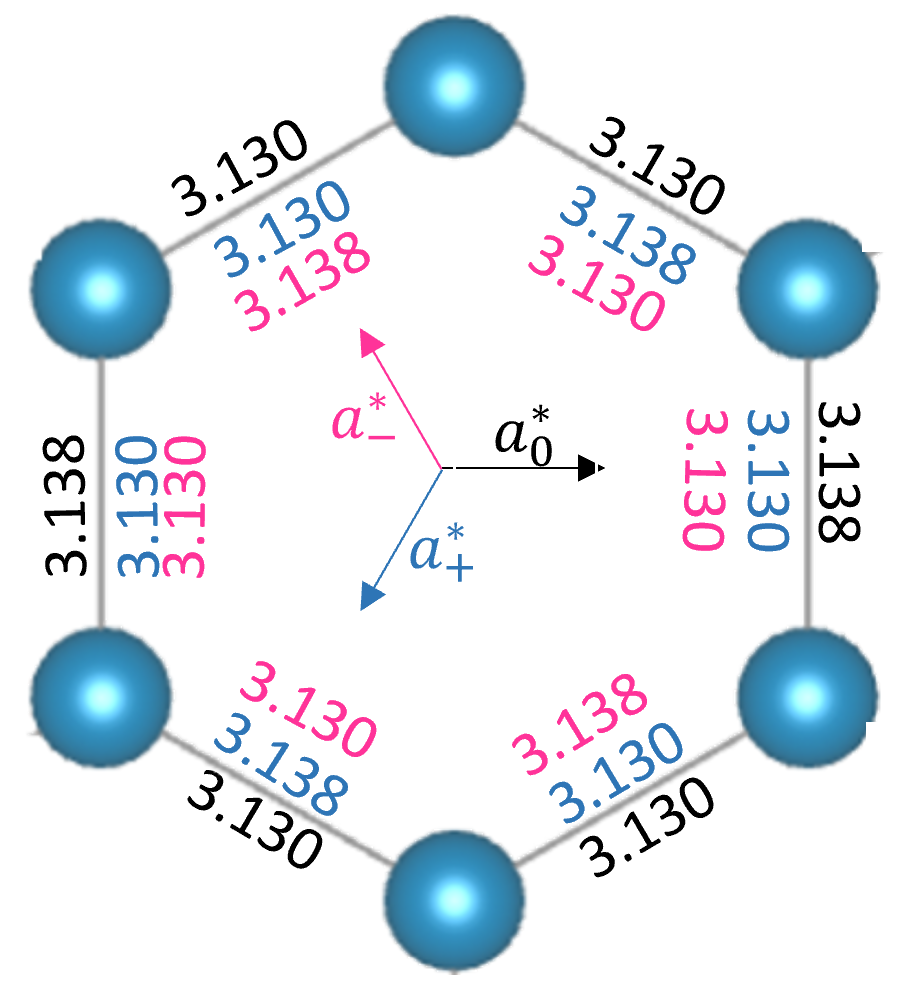}}
\caption{\label{structure} Crystal structure of Na$_2$IrO$_3$. Na atoms are colored yellow, Ir atoms blue and O atoms red. (a) shows the layered structure perpendicular to the $c$-axis which consists of alternating Na and NaIr$_2$O$_6$ layers. The honeycomb structure within the $ab$-plane is depicted in (b).  (c) depicts the honeycomb structure for three orientations differing by rotation of $\pm 120^\circ$ $a^*_0$ (black), $a^*_+$ (blue) and $a^*_-$ (magenta). Due to the slight difference in Ir-Ir-distances (given in \AA~\cite{choi_spin_2012}), there is no perfect overlap of the Ir atoms for the three orientations.}
\end{figure}

As the crystal structure is relevant for the following discussion of our Laue data, we start with a brief overview. Na$_2$IrO$_3$ crystallizes in a monoclinic C2/m structure with one Ir and three Na positions which are listed in Tab.~\ref{latticeparameters}. The lattice constants are $a=5.427(1)$\,\AA, $b=9.395(1)$\,\AA \, and $c=5.614(1)$\,\AA \, ~\cite{singh_antiferromagnetic_2010,moskvin_evidence_2003}. 
Due to its layered structure, which can be seen in Fig.~\ref{structure}(a), consisting of alternating Na and NaIr$_2$O$_6$ layers, Na$_2$IrO$_3$ is prone to stacking faults. 
Within the $ab$-plane, as shown in Fig.~\ref{structure}(b), the Iridium atoms form a honeycomb structure where each Ir atom is surrounded by six Oxygen atoms in octahedral configuration and each honeycomb is filled by one Na atom.

Na$_2$IrO$_3$ crystals grow as thin platelets with the $c^*$-axis perpendicular to the $a$-$b$-plane surface. The honeycomb structure displayed in Fig.~\ref{structure}(c) has almost perfect threefold rotational symmetry around the $c^*$-axis. It is slightly distorted due to trigonal distortions of the individual IrO$_6$ octahedra which change the Ir-O-Ir bond angles from the ideal value of 90$^\circ$ to somewhere in the range of 98-99.4$^\circ$~\cite{choi_spin_2012}. This results in a variation of the Ir-Ir distances ranging from 3.130 to 3.138\,\AA.

\begin{table} [h]

\begin{ruledtabular}
\begin{tabular}{lccccc}

atom & site & $x$ & $y$ & $z$ & $U$\AA$^2$)  \\
\hline
Ir & 4$g$ & 0.5 & 0.167(1) & 0 & 0.001(1) \\
Na1 & 2$a$ & 0 & 0 & 0 & 0.001(6) \\
Na2 & 2$d$ & 0.5 & 0 & 0.5 & 0.009(7) \\
Na3 & 4$h$ & 0.5 & 0.340(2) & 0.5 & 0.009(6) \\
O1 & 8$j$ & 0.748(6) & 0.178(2) & 0.789(6) & 0.001(6) \\
O2 & 4$i$ & 0.711(7) & 0 & 0.204(7) & 0.001(7)

\end{tabular}
\end{ruledtabular}
\caption{\label{latticeparameters} Structural parameters of Na$_2$IrO$_3$ obtained from single-crystal x-ray data at 300\,K~\cite{choi_spin_2012}.} 
\end{table}

\begin{figure}[ht]
\includegraphics[width=0.5\linewidth]{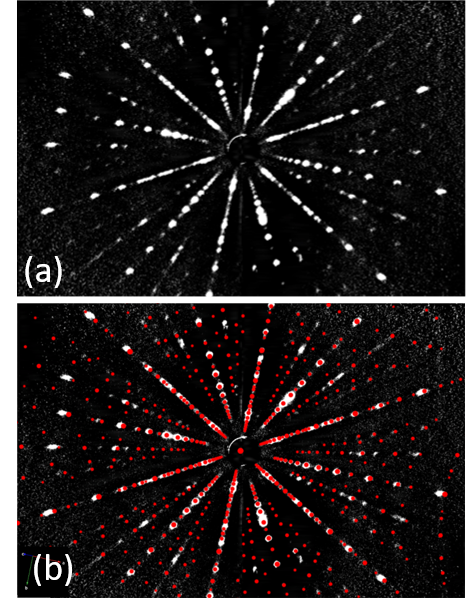}
\caption{\label{Laue_1} (a) Laue pattern of Na$_2$IrO$_3$ recorded with an acceleration voltage of 20kV after background subtraction. (b) The simulated pattern is shown for one of three possible orientations ($a^*_0$).}
\end{figure}

Fig.~\ref{Laue_1}(a) displays a representative measured Laue pattern of one of our crystals. After background subtraction it is first analyzed using the simulation program Single Crystal. The pattern matches reasonably well all three different orientations with mutual $\pm 120^\circ$ rotations around the $c^*$-axis, one being displayed in panel (b). These three orientations shall from now on be referred to as $a^*_0$ ($0^\circ$), $a^*_+$ (rotation of +120$^\circ$) and $a^*_-$ (rotation of  $-120^\circ$). Thus, at first glance, there appears to be no obvious preferred orientation. 

\begin{figure}[ht]
\includegraphics[width=1\linewidth]{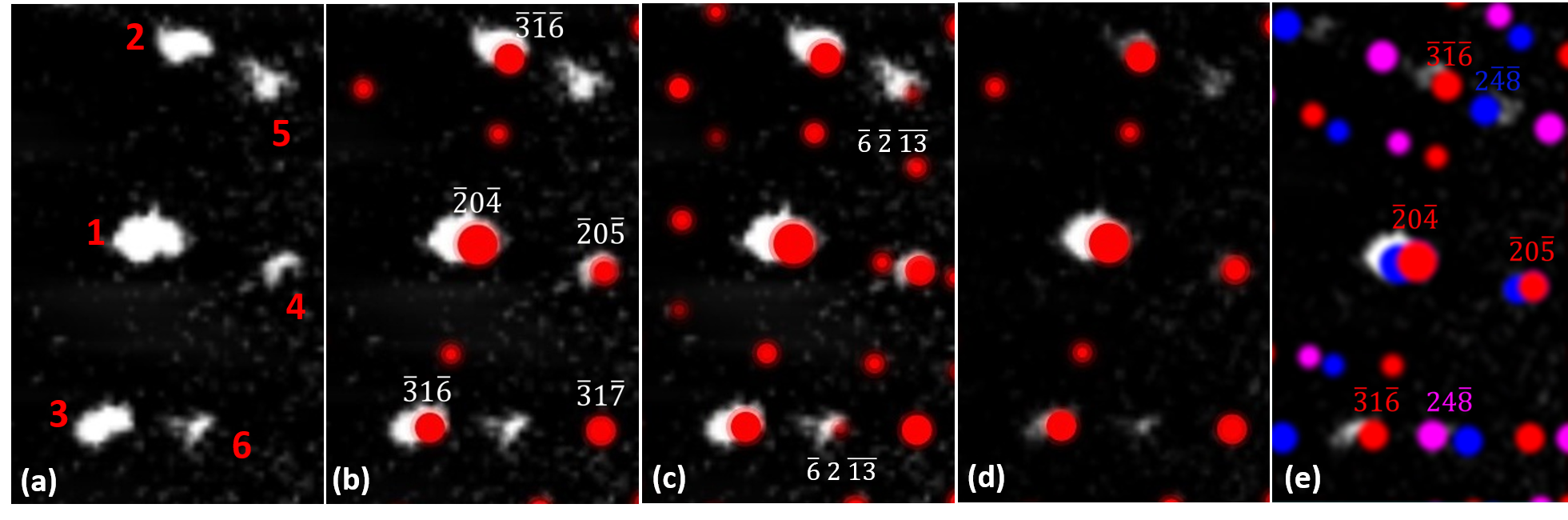}
\caption{\label{Laue_2} Zoomed-in section of the previously depicted Laue pattern recorded at 20kV (a). The relevant reflections are labeled 1-6 in order to be able to distinguish them in the following discussion. (b) shows the simulated reflections (red dots) for one direction ($a^*_0$), (c) those for a simulation with higher detector sensitivity. Here, all detected reflections can be matched, but the simulation also shows several additional signatures which were not detected in the measurement. (d) is the Laue pattern recorded at 15~kV along with the corresponding simulation. Both reflections 5 and 6 are still visible in the measurement yet have disappeared in the simulation. The simulation for simultaneous directions $a^*_0$, $a^*_+$ and $a^*_-$ (red, blue and magenta) in (e) shows a match for all reflections.}
\end{figure}

Taking a closer look at the reflections 1-6 depicted in Fig.~\ref{Laue_2}(a), it becomes apparent that not all reflections can be described by the simulation of just one of the three orientations shown in Fig.~\ref{Laue_2}(b). Here, neither reflection 5 nor 6 is matched by the simulation. One possible explanation for this would be that the detector sensitivity was underestimated. Indeed, a higher sensitivity yields a match for all observed reflections, as shown in Fig.~\ref{Laue_2}(c), where the previously unexplained reflections can now be indexed as  $\overline{6}\,\overline{2}\,\overline{13}$ and $\overline{6}\,2\,\overline{13}$. However, several additional reflections appear in the simulation many of which are not detectable in the recorded Laue pattern. 
Furthermore, while the simulated reflections  disappear for a lower acceleration voltage of 15\,kV, both reflections 5 and 6 remain visible in the corresponding Laue pattern (see Fig.~\ref{Laue_2}(d)). Thus, another explanation is needed. 

Rotating the honeycomb around 120$^\circ$ in either direction (blue or magenta in Fig.~\ref{structure}(c)) leads in some of the overlapping Ir-Ir-bonds to a mismatch in length. Looking at the individual Laue patterns for all three directions, this distortion 
results in most simulated reflections along (h\,0\,l) overlapping for $a^*_0$, $a^*_+$ and $a^*_-$. However, this is not the case for the (0\,k\,l) direction. This leads to clearly visible reflections along (h\,0\,l) whereas those for (0\,k\,l) appear smeared. Reflections where $a^*_0$, $a^*_+$ and $a^*_-$ overlap would show comparatively higher intensities than expected if only one orientation would be present. Indeed, the simultaneous existence of $a^*_0$, $a^*_+$ and $a^*_-$ offers the best match for the recorded data so far as can be seen in Fig.~\ref{Laue_2}(e). Here, reflection 5 is explained by the 2\,$\overline{4}$\,$\overline{8}$ reflection of $a^*_+$ and reflection 6 by the 2\,4\,$\overline{8}$ reflection of $a^*_-$. For signature 1 the reflections for all directions overlap, which explains the increased brightness of this reflection.

Looking again at the reflections $\overline{3} \overline{1} \overline{6}$, $\overline{3} 1 \overline{6}$, $2 \overline{4} \overline{8}$ and  $2 4 \overline{8}$, we find all of them disappear for an acceleration voltage of 10\,kV (not displayed). For 15\,kV, however, the reflections are still detectable. The wavelength of these reflections is close to that for the L$_{\beta_1}$ and L$_{\beta_2}$ characteristic lines of tungsten \cite{ITC} which can be detected for voltages above 12.1\,kV (L$_I$ absorption edge) and thus are absent at lower voltages. 

Furthermore, as the intensity for the characteristic lines is far higher than that of the bremsspectrum, reflections in the immediate vicinity of those lines might also be visible, while other reflections disappear in the background noise. 

\begin{figure}[ht]
\includegraphics[width=0.8\linewidth]{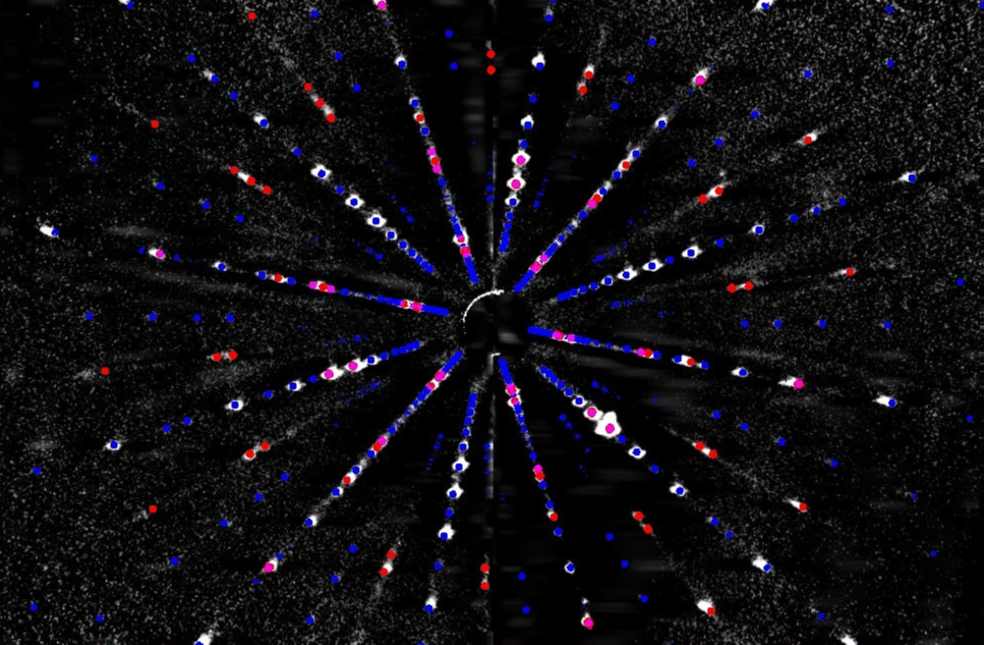}
\caption{\label{Laue_final}  Simulation of the Laue pattern of Na$_2$IrO$_3$ showing reflections visible due to additive overlap in blue, due to proximity to the characteristic lines of tungsten in red and reflections where both occur in magenta. The simulation data was adjusted separately for both detector halves to allow for an optimal result.}
\end{figure}

As the software Single Crystal does not have the capacity for pattern simulation taking into account the above mentioned issues, a self written program considering the characteristic lines as well as the additive overlap of individual reflections was utilized for simulation of such a Laue pattern. Here, we assume a direct proportionality of the intensity to the area of the reflections. Only the characteristic lines for  L$_{\alpha_1}$, L$_{\beta_1}$ and L$_{\beta_2}$ are taken into consideration, as we found those to be the only ones significantly contributing to the diffraction pattern. 
Furthermore, only those reflections for which the wavelength deviates less than  $\Delta\lambda=0.02\,\mathrm{\AA}$ were assigned to the characteristic lines. 
Looking at the resulting Laue pattern, shown in Fig.~\ref{Laue_final}, where both additive overlap and proximity to the characteristic lines of tungsten are taken into consideration for the simulation, we find this to be a good match for all detected reflections. 
Additionally, here the analysis was done separately for both detector halves in order to correct the slight mismatch of both pattern halves stemming from the detector halves not being aligned perfectly.

For Li$_2$IrO$_3$, another member of the honeycomb iridates, stacking faults are attributed to occasional in-plane shifts of the layers by $\pm\vec{b}/3$~\cite{choi_spin_2012, freund_single_2016}. While, however, for a domain containing only a small number of layers, a rotation by 120$^\circ$ yields results similar to those for a translation by $\pm\vec{b}/3$, for larger domains the results for translation and rotation deviate.
Thus, to the best of our knowledge, attributing the Laue reflections to the simultaneous presence of three crystal orientations with 120$^\circ$ rotation offers the best explanation, not only explaining the position of reflections but also the increased brightness of some reflections as well as the smearing of those reflections attributed to the (0\,k\,l) direction.

\subsection{Imaging domain walls by EDX}

\begin{figure}[ht]
\includegraphics[width=0.8\linewidth]{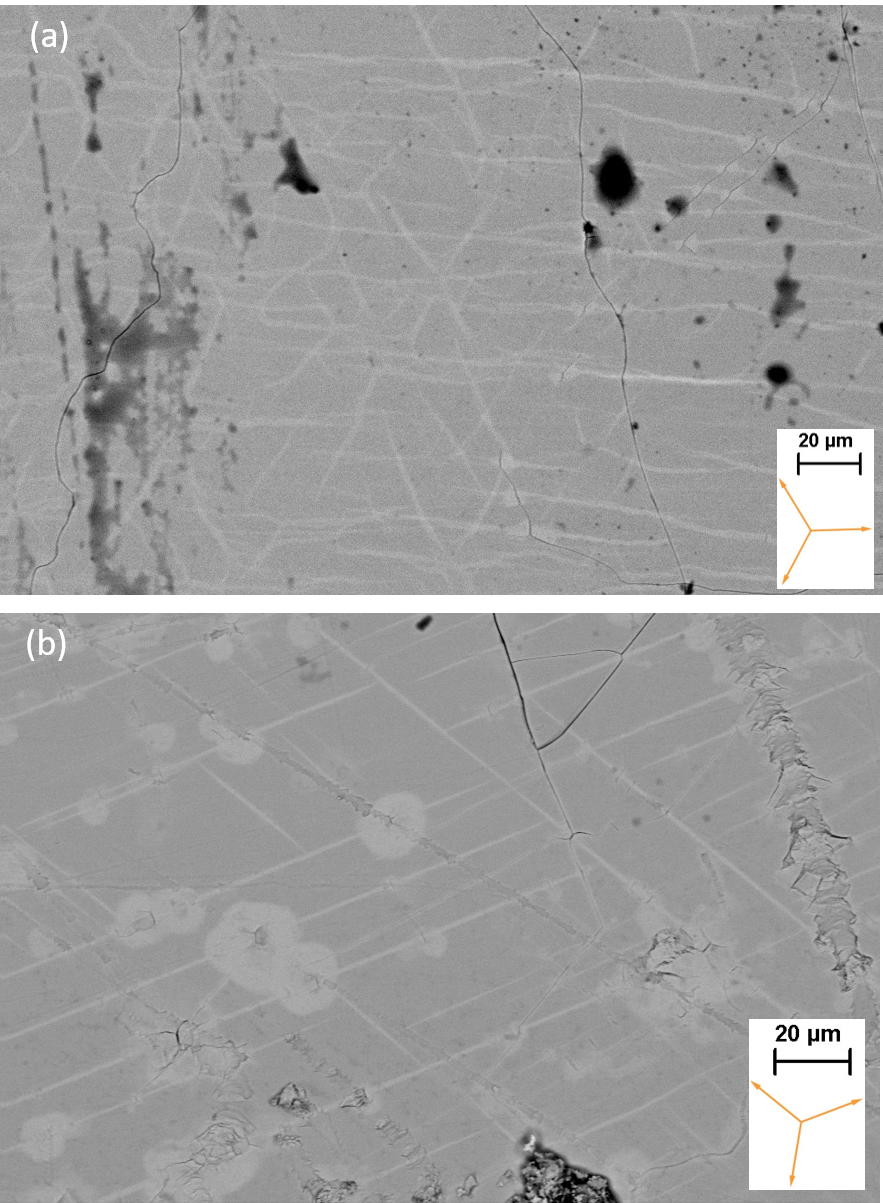}
\caption{\label{SEM_1} SEM image of the surface of two samples of Na$_2$IrO$_3$ recorded using a material contrast sensitive detector. The insets on the right shows the orientation of the $b$-axes. While sample 5 (a) shows a more disordered arrangement, the lines in sample 4 (b) are straight and parallel. Despite these obvious differences, the Na deficient lines appear to be roughly oriented along the $b$-axes determined by Laue analysis for both samples. As Laue measurements were performed before EDX, the depicted samples have been exposed to air for several hours at this point. The dark spots in (a) as well as the bright dots in (b) are likely due to some contamination on the surface or small cracks which can create charging effects appearing bright in the image. This was also observed on other samples previously exposed to air, but is not found on fresh surfaces.}
\end{figure}

\begin{figure}[ht]
\includegraphics[width=0.8\linewidth]{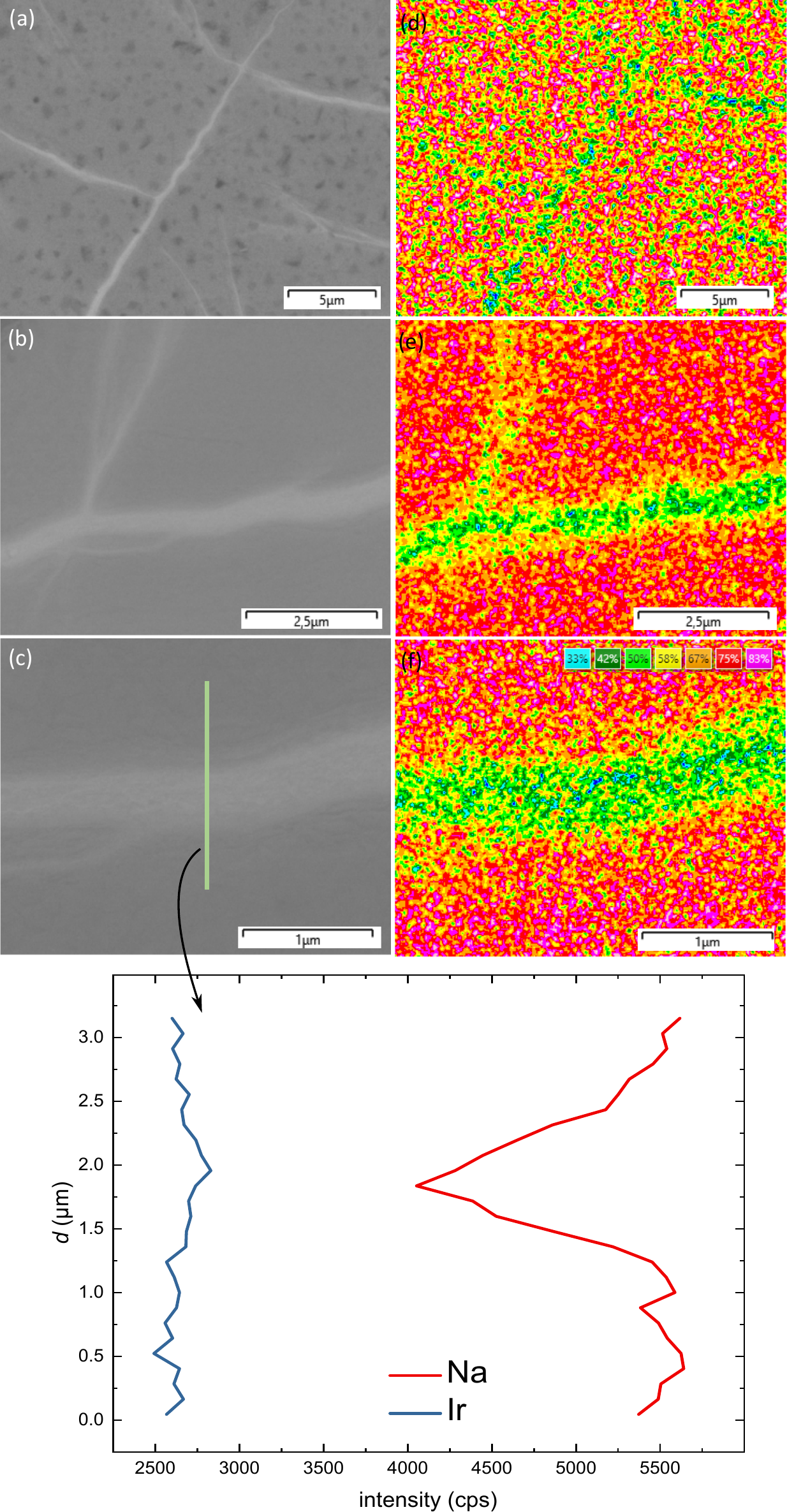}
\caption{\label{EDX_Streifen_1} Elemental distribution of Na determined via EDX for a single line using different magnifications on sample 1. The mapping clearly shows the occurrence of Na deficiency in the lines appearing bright in the SEM image. Here green areas indicate a Na percentage which is diminished compared to the red areas. A line scan of (c) is depicted on the right side to give a quantification of the Na deficiency.}
\end{figure}

SEM images recorded using a material contrast sensitive NTS-BS (Nano Technology Systems - Backscattered Electrons) detector show a series of brighter lines on the surface with a width in the range of up to $2~\mu$m, as shown in Fig.~\ref{SEM_1}.
These lines can be observed on untreated or polished crystals as well as on freshly cleaved surfaces and could also be found using other detectors, such as HE-SE2 (High Efficiency Secondary Electron) or AsB (Angular selective Backscattered electron)), where they appeared, however, far less prominent. While some of the lines are arranged parallel to each other, others are rotated by either 60$^\circ$ or 120$^\circ$. Comparing the lines to the orientation of the crystal previously determined by Laue, many lines seem to be running along at least one of the three $b$-axes.  While some samples show more irregularly distributed and overall less straight lines, a rough assignment to the $b$-axes remains possible for all investigated crystals for at least one direction. Through thorough investigation of the surface, we exclude topographical features such as steps, cracks or depositions as the origin of these lines. Combining SEM and Laue data we thus propose to assign these lines to structural domain walls separating differently oriented domains of Na$_2$IrO$_3$.  Note that in case of (Na$_{1-x}$Li$_{x})_{2}$IrO$_{3}$ for $x>0.25$ in the phase separated regime, also line structures were found in EDX, which however, can be associated with Li$_2$IrO$_3$~\cite{Manni2014}.

Fig.~\ref{EDX_Streifen_1} shows an enlarged image of these lines along with elemental distribution maps for sodium. Here green areas show a lower Na concentration compared to red ones. A linescan taken perpendicular to one of the lines clearly shows a dip in Na concentration which coincides with the brighter region in the SEM image.

STM measurements previously showed Na$_2$IrO$_3$ to be Na deficient on the surface with up to 2/3 of the Na atoms missing~\cite{lupke2015}. However, this method is limited to the surface, whereas EDX can probe a volume of several hundred nanometers into the crystal \cite{castaing1960electron}. All EDX data was recorded using a voltage of 20\,keV. Tab.~\ref{NaIr} summarizes the Na to Ir intensity ratios on the lines (i.e. on the domain walls), as well as the surrounding area for 6 different samples. First, we note that even away from the lines the ratio differs from the stoichiometric value of 2. A reduction of the intensity ratio below 2 even away from domain boundaries on freshly cleaved samples likely points to the occurrence of re-absorption of the heavy Iridium atoms during EDX. Therefore, the intensity ratio does not directly equal the atomic ratio, though it is a measure of it. Clearly it is drastically reduced on the lines compared to the surrounding regions. The Na deficiency effectively hole-dopes the iridate layers, whose impact on electrical transport is addressed below.


\begin{table} [h]

\begin{ruledtabular}
\begin{tabular}{lccc}

sample & (Na/Ir)$_\mathrm{line}$ & (Na/Ir)$_\mathrm{s}$ \\
\hline
1 & 1.5(2) & 1.8(1)  \\
2 & 1.5(2) & 1.8(1)  \\
3 & 1.3(1)  & 1.8(1) \\
4 & 1.3(1)  & 1.7(1) \\
5 & 1.2(2) & 1.6(2) \\
6 & 1.1(2) & 1.6(2)\\

%

\end{tabular}
\end{ruledtabular}
\caption{\label{NaIr} Na/Ir ratio determined by EDX for several samples at the detected lines and on the surrounding area (s).} 
\end{table}

The lower Na concentration along the structural domain walls also explains their bright appearance in the SEM image, as within them the average atomic number is increased. 
It is not yet clear how deep into the crystal the lines penetrate. The fact that upon cleaving a  crystal, lines could instantly be found on the fresh surfaces implies either an instantaneous formation upon exposition or a penetration depth reaching far into the crystal. 

\subsection{Electrical resistivity of domain walls}


\begin{figure}[ht]
\includegraphics[width=0.8\linewidth]{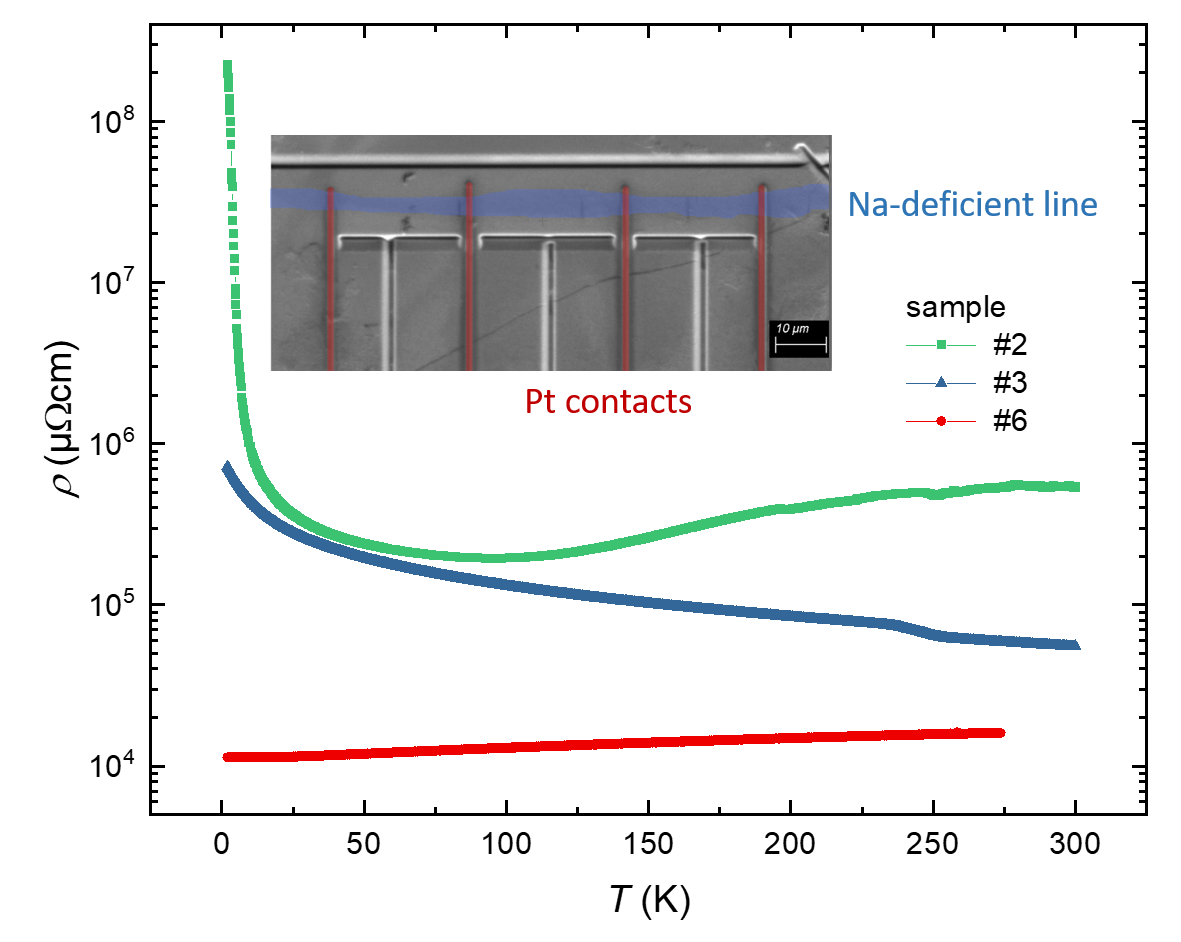} 
\caption{\label{Streifen1} Temperature dependence of the electrical transport of Na deficient lines on the surface of Na$_2$IrO$_3$ crystals with varying Na content. Decreasing Na concentration lowers the resistance due to hole doping. The sample with lowest Na content is completely metallic (red line). The inset shows a microstructure prepared and electrically contacted by FIB.}
\end{figure}  

\begin{figure}[ht]
\includegraphics[width=0.8\linewidth]{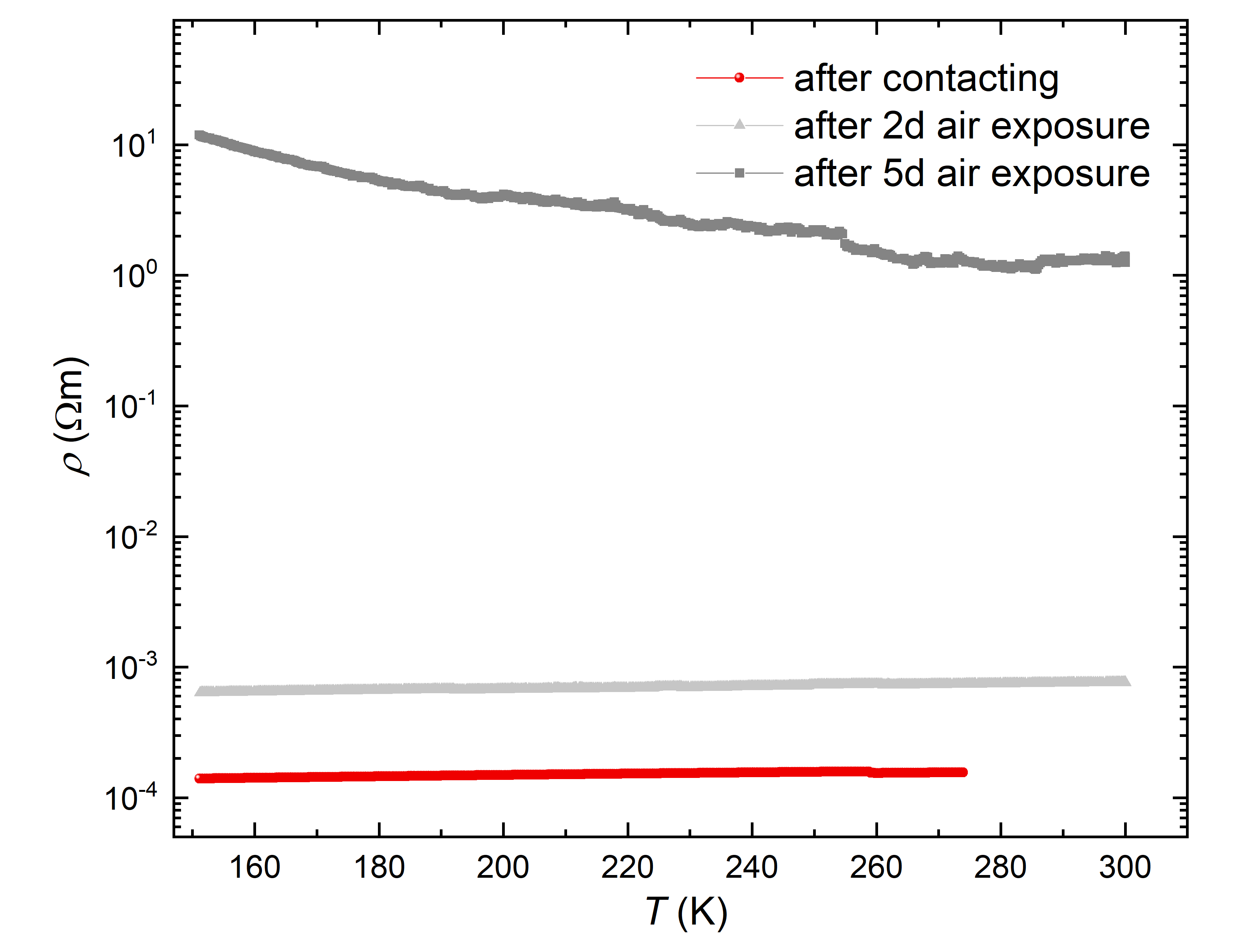} 
\caption{\label{Streifen_deg} Temperature dependence of the electrical resistance along a structural domain wall in sample 6 measured directly after contacting (red curve, from Fig.\ref{Streifen1}), as well as after 2 days (light gray) and 5 days (gray) of air exposure.}
\end{figure}  

\begin{figure}[ht]
\includegraphics[width=0.7\linewidth]{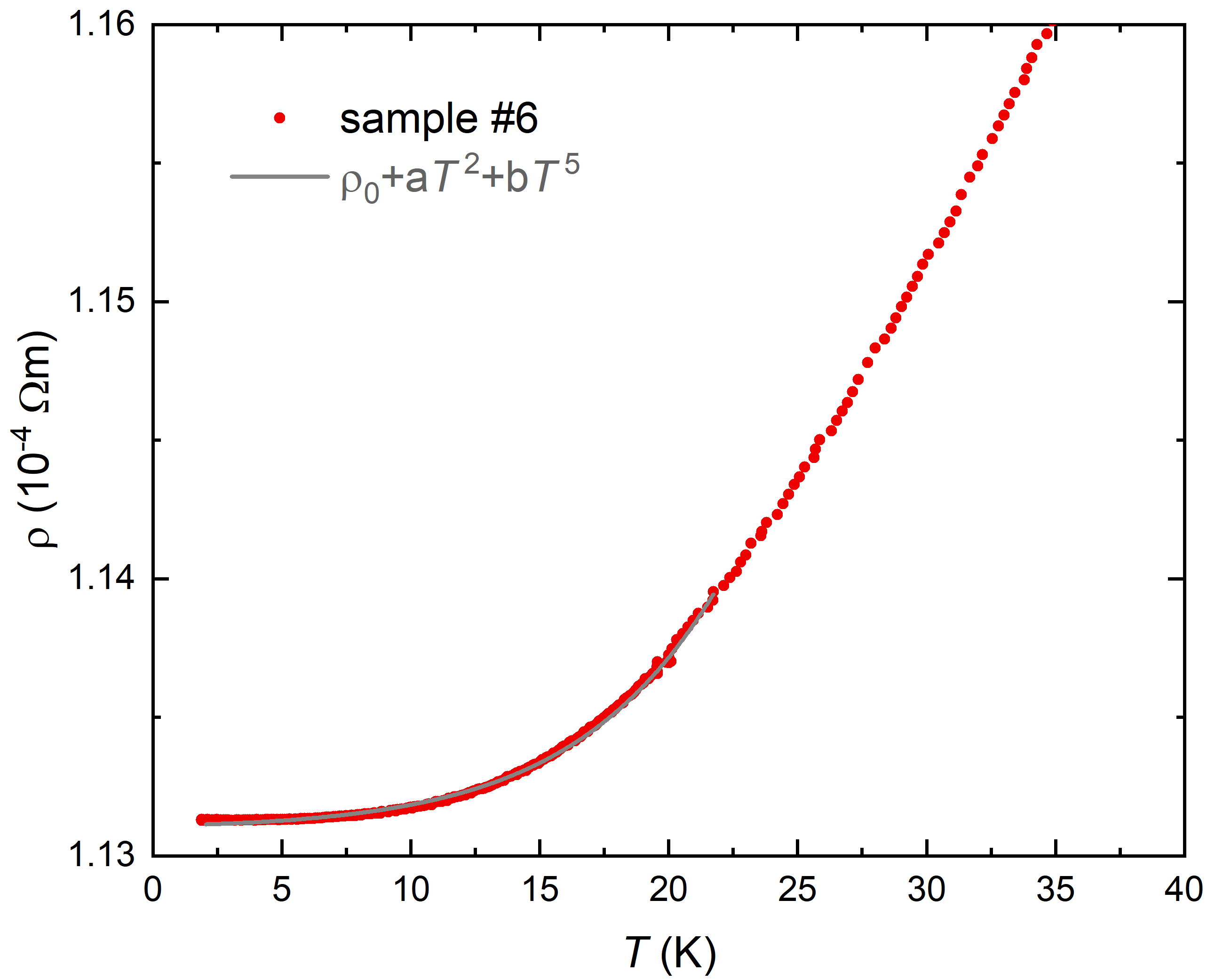} 
\caption{\label{lowT} Low temperature electrical resistivity along a structural domain wall in sample 6. The solid line represents correlated metallic behavior according to $\rho(T)=\rho_0+aT^2+bT^5$ with $\rho_0=1.13\cdot 10^{4} \mu \Omega$cm, $a=6.18 \cdot 10^{-2}~\mu \Omega$cm/K$^2$ and $b=1.11\cdot 10^{-5}~\mu\Omega$cm/K$^5$.}
\end{figure}

In order to study the electrical resistance of Na-deficient structural domain walls, FIB-assisted microsectioning and electrical contacting was utilized. Crystals were glued onto sapphire substrates using epoxy resin, cleaved inside an Argon filled glovebox, transferred to a sputter chamber without air exposure and partially coated with gold. The samples were then structured using a FIB-SEM in a way that allowed for a four probe measurement along a defined area by contacting a line using gas-injection system (GIS) deposited Pt within the FIB chamber and cutting trenches to define the desired geometry \cite{moll2018focused}, as can be seen in the inset of Fig.~\ref{Streifen1}. Contacting was done in the same way for all depicted samples. Air contact of the gold covered sample surface was limited to the transfer from the sputter chamber to the FIB.  EDX was used to determine the Na-Ir ratio along the lines compared to the surrounding surface. Resistivity data for three such lines, each of which was recorded on a different crystal, can be seen in Fig.~\ref{Streifen1}. 

The absolute resistance was scaled with the geometrical parameters, such as the width of the lines from SEM, the distance of the voltage contacts, as well as the total sample thickness, to obtain the resistivity. This gives us an upper bound of the resistivity, since less Na-deficiency in the inner parts, deeper below the surface, may reduce the effective thickness of the conducting regions. While absolute values are already rather small compared to previous measurements, we observe a further decrease for lower Na content of the lines. Sample 6 shows metallic behavior ($d\rho/dT>0$) down to lowest temperatures. To check, whether the metallic behavior is preserved over time, we re-measured the electrical resistance of sample 6 two subsequent times after air exposure. Prolonged contact with CO$_2$ and H$_2$O is known to lead to sample degradation \cite{krizan2014}. As shown in Fig. \ref{Streifen_deg}, after two days of exposure to air, the resistivity still appeared metallic, yet the absolute values increased. After five days the resistivity shows entirely insulating behavior. The metallic behavior of sample 6 before air exposure cannot be explained by degradation, which by contrast enhances the resistivity. Instead, the metallic resistance indicates that the metal-insulator transition has been passed for Na deficient, i.e. hole-doped Na$_2$IrO$_3$.

Finally, we analyze the low-temperature data, cf. Fig. \ref{lowT}. The temperature dependence can be described using a sum of the residual resistance, electron–electron and electron–phonon scattering contributions according to $\rho=\rho_0+aT^2+bT^5$ \cite{ashcroft_solid_1976}. The fitted coefficient $a=0.062 \mu\Omega$cm/K$^2$ lies in between values for bilayer strontium iridate Sr$_2$RuO$_4$ along the c- and a-axis and it is comparable to values for moderately heavy 4f- and 5f- Kondo systems~\cite{maeno1997}. But of course this relies on the above assumption of a homogeneous Na deficiency across the entire sample thickness of $\sim 20-40$\,$\mu$m.

\section{{Conclusion}}

In conclusion, Laue pattern of Na$_2$IrO$_3$ indicate structural domains and are perfectly well described by additive overlap of reflections from three different structural twins with 120$^\circ$ rotated $b$-axes around the $c^*$-axis. Complementary scanning electron microscopy (SEM) energy dispersive x-ray analysis on cleaved crystal surfaces revealed lines roughly along the different b-axes with significant (up to 1/3) Na deficiency compared to the surrounding areas. We thus associate these lines to structural domain walls. Resistivity measurements along individual domain walls in FIB-assisted microstructured samples reveal a dependence of the resistance on their Na to Ir ratio, with lower Na content and thus increased hole doping corresponding to lower resistivity. For the domain wall with highest Na deficiency a completely metallic behavior down to 2~K has been found, indicating that Na$_2$IrO$_3$  can be tuned locally across the metal-insulator transition. The observed coefficient of $T^2$ resistivity behavior suggests a moderately correlated metal state, though the absolute values of the resistance are governed by the uncertainty of the effective thickness of the domain boundaries.

At first glance, this behavior reminds to conductive domain walls observed in other otherwise insulating materials, such as ferroelectrics \cite{seidel_conduction_2009, maksymovych_dynamic_2011, ghara_giant_2021} where conductivity can be tuned by the application of an electric field. This leads to non-Ohmic I-V-characteristics which have also been observed for Na$_2$IrO$_3$ \cite{rodriguez_competition_2020, dziuba_combined_2022}. However, in the case of Na$_2$IrO$_3$ it is known from ARPES measurements that Na deficiency or additional deposited Na atoms influence the charge carrier density and thus the Fermi level \cite{alidoust_observation_2016, moreschini_quasiparticles_2017, rodriguez_competition_2020} and our analysis reveals a clear Na deficiency at the domain walls, which is thus at the origin of the conducting behavior.


\section*{Acknowledgements}

Technical support by Klaus Wiedenmann is gratefully acknowledged. We thank Dr. Toni Helm for advise regarding the FIB preparation of samples. This work was supported by the German Research Foundation project 492547816 (TRR 360).


\bibliography{Bib_NIO}

\end{document}